\documentclass[12pt]{iopart}
\usepackage{graphicx}
\usepackage{umlaut}
\usepackage[english]{babel}

\begin{document}
\title{Graphene on insulating crystalline substrates}

\author{S. Akcöltekin, M. El Kharrazi, B. Köhler, A. Lorke, M. Schleberger \footnote{electronic address: marika.schleberger@uni-due.de}} 
\address{Fachbereich Physik, CeNIDE, Universität Duisburg-Essen, D-47048 Duisburg, Germany}

\begin{abstract}
We show that it is possible to prepare and identify ultra--thin sheets of graphene on crystalline substrates such as SrTiO$_3$, TiO$_2$, Al$_2$O$_3$ and CaF$_2$ by standard techniques (mechanical exfoliation, optical and atomic force microscopy). On the substrates under consideration we find a similar distribution of single, bi- and few layer graphene and graphite flakes as with conventional SiO$_2$ substrates. The optical contrast $C$ of a single graphene layer on any of those substrates is determined by calculating the optical properties of a two-dimensional metallic sheet on the surface of a dielectric, which yields values between $C=$~-1.5\% (G/TiO$_2$) and $C=$~-8.8\% (G/CaF$_2$). This contrast is in reasonable agreement with experimental data and is sufficient to make identification by an optical microscope possible. The graphene layers cover the crystalline substrate in a carpet-like mode and the height of single layer graphene on any of the crystalline substrates as determined by atomic force microscopy is $d_{SLG}=0.34$~nm and thus much smaller than on SiO$_2$. 
\end{abstract}
%

\maketitle
\section{Introduction}
Graphene has recently attracted much attention due to its unique electronic properties (see e.~g.~ \cite{Novoselov05,Geim07}). Part of graphene's attraction comes from the fact that it is fairly easy to produce. The easy preparation of graphene is not only due to the simple mechanical exfoliation process but in part also due to the Si-SiO$_2$ substrate it is usually prepared on, which offers several advantages. It is relatively cheap and easy to obtain, easy to clean, and with an appropriately chosen SiO$_2$ thickness it is very easy to identify even single layer thick sheets of graphene. As a consequence, the preparation of graphene on SiO$_2$ has been perfected by many groups and a variety of important results has been obtained from this type of samples. 

However, there are a number of reasons to study graphene on a larger variety of substrates. For example, any effects connected to the type of substrate like doping, binding properties or crystallinity could be studied and possibly exploited. The current focus on SiO$_2$ is not so much given by any unique physical property but simply by the fact that graphene sits on a {\em thin} dielectric, which makes it easy to identify, even with standard optical microscopy. It was shown that the astonishing visibility of graphene is mainly due to an interference effect between the graphene and the thin layer of SiO$_2$ \cite{Ni07,Blake07}. The optimum SiO$_2$ thickness to make graphene visible was calculated using the Fresnel formula to be 90~nm. This results in a maximum contrast of $C \approx$ 12\% which is easily visible in a standard optical microscope. In the following we demonstrate that an appreciable contrast can be achieved on bulk dielectrics without the need of thin films. This opens up the possibility to study graphene on a large variety of substrates, which may be crystalline and have properties and interactions with graphene, which greatly differ from those of SiO$_2$.

\section{Experimental details}
\begin{figure}[htb]
\begin{center}
\includegraphics[width=10.0cm]{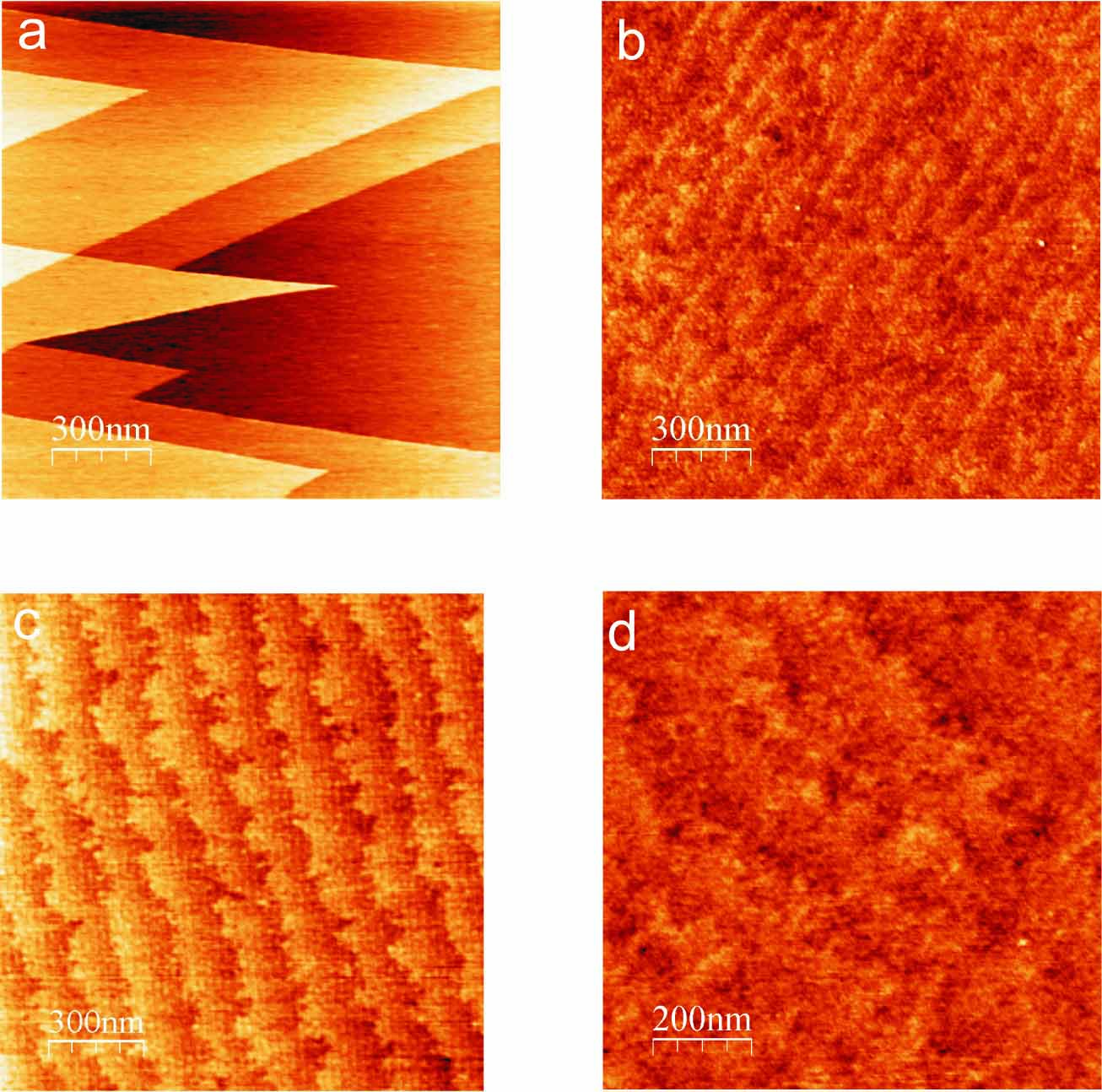}
\caption{Image taken by AFM (Nanosensors PPP-NCHR, $f_0=209-497$~kHz) of the pristine substrate surfaces. a) CaF$_2$, b)  Al$_2$O$_3$, c) SrTiO$_3$, d) TiO$_2$. Terraces are separated by monoatomic straight (CaF$_2$) or fuzzy (oxides) step edges. Image size indicated by scale bars.}
\label{figure1}
\end{center}
\end{figure}

We applied the standard exfoliation technique \cite{Geim07} to deposit graphene on SrTiO$_3$(100), TiO$_2$(100), Al$_2$O$_3$(1102) and CaF$_2$(111) single crystal surfaces (Crystec, Korth). These materials are widely used and are commercially available with very high surface quality. Fig.~\ref{figure1} shows atomic force microscope (AFM DI-3100, intermittent contact mode) images of clean samples taken directly from the package without any further surface treatment. Terraces are easily identified, the step edges however appear fuzzy, except in the case of CaF$_2$ where straight cleavage steps are typical. The mean roughness is below 0.5~nm and usually no contaminations are detected on the oxidic surfaces, whereas CaF$_2$ is easily contaminated by adsorbants, most likely water \cite{SAkcoeltekin09}. For optical microscopy (see fig.~\ref{figure2}) we used an Olympus BX40 (100x NA 0.9) and Raman spectra to determine the number of graphene layers were recorded with a Jobin Yvon--LabRam microscope system. The spectra were excited with a laser of 633~nm wavelength and a laser power of $\approx 0.4$~mW. Single layer and bilayer flakes were identified via the shape of the $2D$ line \cite{Ferrari06}.

\section{Results and Discussion}
\begin{figure}[htb]
\begin{center}
\includegraphics[width=15cm]{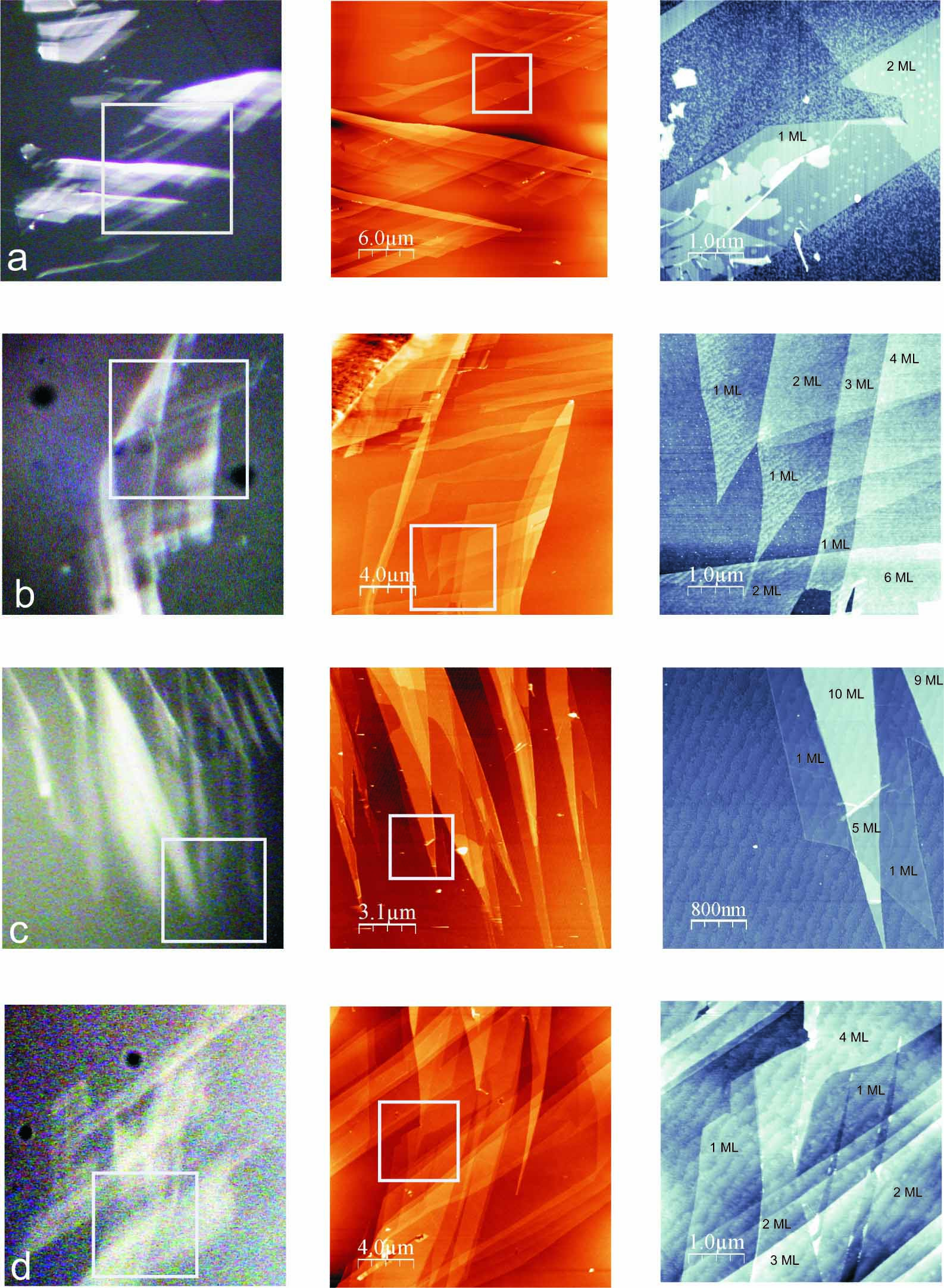}
\caption{a) CaF$_2$, b)  Al$_2$O$_3$, c) SrTiO$_3$, d) TiO$_2$. Each panel from left to right: Images taken by optical microscopy (left). AFM (middle and right) of graphene crystallites consisting of different numbers of graphene sheets found on the surface region marked by the square. Size of AFM images indicated by scale bars.}
\label{figure2}
\end{center}
\end{figure}

When sheets of graphene from a single crystal of highly oriented pyrolitic graphite are exfoliated onto a pristine surface of any of those crystals, results as shown in fig.~\ref{figure2} are obtained. Typically, flakes the size of several square microns stick to the substrate. Most of the flakes are identified as thick sheets of graphite (G), but we also find graphene with a thickness of a few monolayers (FLG) or eventually  single layers (SLG). This is comparable with what we find on substrates with 90~nm SiO$_2$. In the case of CaF$_2$ adsorbates are trapped underneath the graphene sheet, which influence the contrast and show up as bright patches in the right panel of Fig. \ref{figure2}(a). The optical contrast compared to graphene on 90~nm SiO$_2$ is rather poor on all substrates (a few per cent instead of 12\%, see table~\ref{table1}) with the exception of CaF$_2$. But in all cases it is still sufficient to identify regions with single layer flakes. 

The clear visibility of even single graphene layers demonstrated in Fig.~\ref{figure2}, is somewhat surprising in view of the general assumption that an appropriately chosen layer of SiO$_2$ is a prerequisite for good optical contrast. In the following, we will discuss in more detail this finding, which will open up a large class of possible substrates for the investigation of graphene. Abergel et al.~already pointed out that in principle a thin film is not necessary to make graphene visible as long as a resonance condition with the substrate is fulfilled \cite{Abergel07}. Here, we have made use of another approach to calculate the optical contrast of a thin layer of graphene on any kind of (semi--infinite) substrate without a thin film in between. It is based on a transfer matrix approach to describe the reflection and transmission properties of an electromagnetic wave interacting with stratified media \cite{Born}. For the sake of clarity we assume perpendicular incidence $\theta=0$. Our system is modelled by a semi-infinite substrate (z $<$ 0, index 3), a thin layer of graphene (0 $<$ z $<$ $\Delta$, index 2) and a semi-infinite slab of air (z $>\Delta$, index 1). The transfer matrix associated with the graphene is given by:

\begin{equation}
M(\Delta)=\left(
\begin{array}{cc}
	\cos(k_0 \hat n_2 \Delta) & -i\frac{1}{\sqrt{\hat \varepsilon_2}}\sin(k_0 \hat n_2 \Delta)\\
	-i\sqrt{\hat \varepsilon_2}\sin(k_0 \hat n_2 \Delta) & \cos(k_0 \hat n_2 \Delta)\\
\end{array}\right)
\label{eq:1}
\end{equation}
where $k_0$ is the wave vector, $\hat n_2 = \sqrt{\hat\varepsilon_2}$ is the complex refractive index, where the complex dieletric constant is given by the complex electric conductivity: $\hat{\varepsilon}=\frac{i\hat{\sigma}}{\varepsilon_0\, \omega}$. The coefficients for transmission and reflection are then given by:
\begin{equation}r=\frac{(m_{11}+m_{12}\hat n_3)\hat n_1-(m_{21}+m_{22}\hat n_3)}
{(m_{11}+m_{12}\hat n_3)\hat n_1+(m_{21}+m_{22}\hat n_3)}
\label{eq:2}
\end{equation}
\begin{equation}
t=\frac{2\hat n_1}
{(m_{11}+m_{12}\hat n_3)\hat n_1+(m_{21}+m_{22}\hat n_3)}
\label{eq:3}
\end{equation}
with $m_{ij}$ the matrix elements. We are interested in the contrast $C=\frac{R_0-R}{R_0}$ which can be calculated from the reflected intensity with graphene $R=|r|^2$ and the intensity of the reflected light without graphene $R_0=\left|\frac{\hat n_1-\hat n_3}{\hat n_1+\hat n_3}\right|^2$. Thus, we need to determine $r$ by using the matrix elements given by eq.~\ref{eq:1}:
\begin{equation}
r=\frac{(\cos(k_0 \hat n_2\Delta)-i\frac{1}{\sqrt{\hat\varepsilon_2}}\sin(k_0 \hat n_2\Delta)\hat n_3)\hat n_1-(-i\sqrt{\hat\varepsilon_2}\sin(k_0 \hat n_2\Delta)+\cos(k_0 \hat n_2\Delta)\hat n_3)}
{(\cos(k_0 \hat n_2\Delta)-i\frac{1}{\sqrt{\hat\varepsilon_2}}\sin(k_0 \hat n_2\Delta)\hat n_3)\hat n_1+(-i\sqrt{\hat\varepsilon_2}\sin(k_0 \hat n_2\Delta)+\cos(k_0 \hat n_2\Delta) \hat n_3)}
\label{eq:4}
\end{equation}

To evaluate this expression we use the fact that the thickness $\Delta$ of the graphene is orders of magnitude smaller than the wavelength of visible light. Therefore, we treat Eq. (4) in the limit $\Delta \rightarrow 0$, and account for the graphene layer using a two-dimensional metallic sheet of conductivity $\hat \sigma^{2D}$. To calculate the coeffcient $r$ we expand the sine and cosine:
$$\cos(k_0 \hat{n}\Delta)\approx 1-\left(\frac{k_0 \hat{n}\Delta}{2}\right)^2, ~~~~ \sin(k_0 \hat{n}\Delta)\approx k_0 \hat{n}\Delta-\left(\frac{k_0 \hat{n}\Delta}{6}\right)^3$$

With these expansions and in the limit of $\Delta \rightarrow 0$ eq.~\ref{eq:4} becomes:

\begin{equation}
r=\frac{\hat n_1-\hat n_3-\frac{\hat{\sigma}^{2D}}{\varepsilon_0 c}}{\hat n_1+\hat n_3+\frac{\hat{\sigma}^{2D}}{\varepsilon_0 c}}
\label{eq:7}
\end{equation}

If we take the minimum conductivity of a single layer of graphene $\hat{\sigma}^{2D}=\frac{e^2}{4\hbar}$ \cite{Abergel07} and set $\hat n_1$=1 for air we obtain for any kind of substrate: $R=\left|\frac{0.9770 -\hat n_3}{1.0229+ \hat n_3}\right|^2$. The corresponding contrast values that were obtained for the substrates used in the experiments are listed in table~\ref{table1}. As all imaginary parts here are negligible only the real parts of the refractive indices (taken from  \cite{CRC}) were used for the calculation. For comparison we added the values for graphene on a thin film of SiO$_2$ as calculated by Blake {\em et al.} \cite{Blake07}. The minus sign indicates that the flake will appear brighter than the substrate. The values we find in our experiment \cite{IGORPro} are in reasonable agreement with the calculated contrast values. In the case of CaF$_2$, which yields the highest contrast, the lower experimental value might be due to the adsorbates which are present on CaF$_2$ cleaved and imaged under ambient conditions \cite{SAkcoeltekin09}. A lower limit for the optical detection of graphene is in this study represented by TiO$_2$ and SrTiO$_3$ due to their high refractive index yielding contrast values of only $C\approx$~-2\%.

\begin{table}
\begin{center}
\begin{tabular}{|c|c|c|c|}
\hline
Substrate & $n$ at 589~nm & $C_{calc}$ in \% & $C_{exp}$ in \%\\
\hline
\hline
90~nm SiO$_2$ & 1.5442 & 5 -- 12 \cite{Blake07} & +8\\
SrTiO$_3$ & 2.4082 & -1.9 & -2.2 \\
TiO$_2$ & 2.612 & -1.5 & -3.0\\
Al$_2$O$_3$ & 1.7673 & -4.3 & -2.8\\
CaF$_2$ & 1.4338 & -8.8 & -5.9\\
\hline
\end{tabular}
\end{center}
\caption{Calculated and experimentally determined contrast values for various substrates.}
\label{table1}
\end{table}
\vspace{0.5cm}

\begin{figure}[htb]
\begin{center}
\includegraphics[width=12cm]{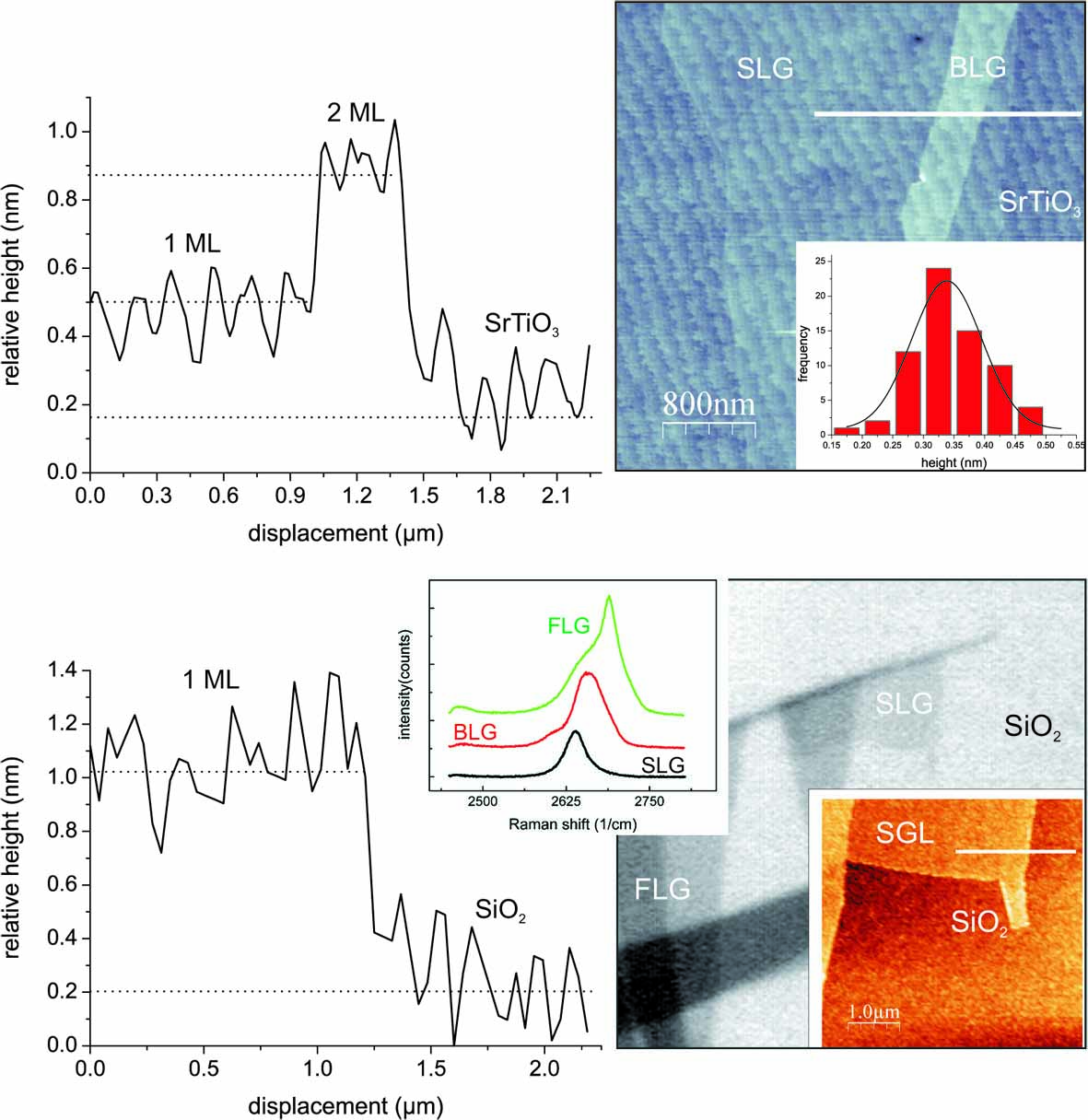}
\caption{Top: representative line scan and height distribution of graphene layers on SrTiO$_3$ (AFM image on the right). Dotted lines represent mean values as determined from height distributions. Bottom: representative line scan to determine height of graphene layers on SiO$_2$ (optical microscopy and AFM on the right). Dotted lines represent mean values. The inset shows Raman spectra taken from the same flake clearly showing spectral features associated with single layer, bilayer and few layer graphene.}
\label{figure3}
\end{center}
\end{figure}

We analyzed the height of single layer ($d_{SLG}$) as well as of bilayer graphene ($d_{BLG}$) on the different substrates by determining the step height from $\geq$~80 individual line scans (see fig.~\ref{figure3}) for each substrate. The resulting height distributions were fitted by a gaussian yielding the mean height and the standard deviation. The baseline was determined by averaging over several line scans taken from the respective uncovered substrate. Raman spectra were recorded to provide a definitive identification of single layer and bilayer flakes independently of AFM. In the case of SiO$_2$, additionally, the interference colours were used to identify single and bilayer flakes. From our combined optical, Raman- and AFM data we find an average height of the first graphene layer on SrTiO$_3$, TiO$_2$ and Al$_2$O$_3$ of $d_{SLG}=(0.34\pm0.01$)~nm. This is a much lower value than on SiO$_2$ with $d_{SLG}=(0.83\pm0.02)$~nm (see fig.~\ref{figure3}), the latter being in good agreement with the literature \cite{Obraztsova08}. The step height between the first and the second layer of graphene is again different on the crystalline substrates compared to SiO$_2$. For the oxides we find for the thickness of the bilayer $d_{BLG}=(0.69\pm0.02)$~nm, which is twice the value of the single layer. On SiO$_2$ the value is $d_{BLG}=(1.2\pm0.4)$~nm, yielding an interlayer distance between the first two layers close to 0.4~nm but afflicted with a large deviation. 

The values on the crystalline substrates of $d_{SLG}=\frac{1}{2}d_{BLG}=0.34$~nm correspond very well to the nominal extention of the unperturbed $\pi$-orbitals of a perfect single graphene sheet. This indicates a very hight quality of the sheets as well as a strong bonding of the graphene to the crystalline substrate. This is an important finding as it has already been discussed that the substrate plays a major role for the transport properties of graphene. The substrate induces charged impurities which lead to a locallly increased density of charge carriers \cite{Adam07,Chen08}. A change in mobilities can thus be achieved by controlling the substrate. This is consistent with the experimental observation of high mobilities in free standing graphene sheets \cite{Du08}. 

\begin{figure}[htb]
\begin{center}
\includegraphics[width=12cm]{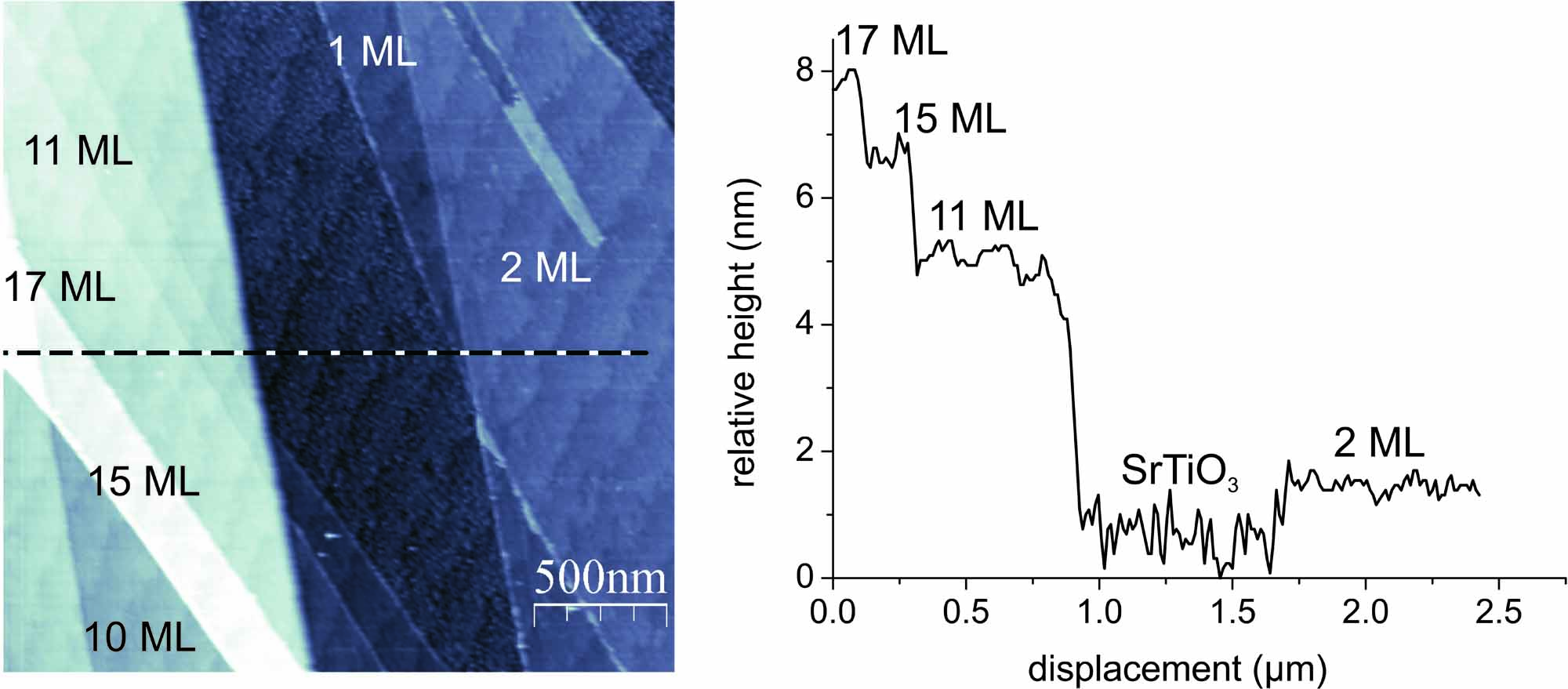} 
\caption{Left: AFM image of thick graphene crystallite on SrTiO$_3$. Step edges can still be seen under 15~ML of graphene. Image size 2.5 $\times$ 2.5 $\mu$m$^2$. Right: Line scan along the black line shown in the AFM image. }
\label{figure4}
\end{center}
\end{figure}

In an earlier study it was shown that to some extent graphene follows the texture of a GaAs substrate \cite{Stoeberl08}. The authors find that the stiffness of the graphene prevents it from following substrate corrugations larger in height than $\approx$ 8 nm in accordance with the values found for graphene on SiO$_2$ \cite{Ishigami07}. From our AFM images it can be seen that the flakes cover the substrate terraces (corrugation less than 0.4~nm) like a carpet as the step edges are still clearly detected by the AFM giving the impression of a transparent veil (see fig.~\ref{figure4}). Note, that the steps are reproduced very well but the small-scale variations on the terraces are not (see line scan in fig.\ref{figure4}). For graphene sheet thicknesses $\geq$ 17~ML the step edges can not be identified anymore in a topographic image. 

\section{Conclusions}
In conclusion, we have demonstrated that the standard techniques for preparing thin graphene sheets on SiO$_2$ can be extended to a variety of dielectrics. This opens up an additional range of useful (affordable, easy to handle, high surface quality) substrates and makes it possible to study how the interaction of graphene with its supporting material affects its electronic, magnetic and optical properties. From the fact that the measured height of single graphene layers on SrTiO$_3$, TiO$_2$ and Al$_2$O$_3$ is only 0.34~nm, we conclude that this interaction is much stronger for these materials than for SiO$_2$, were the corresponding value is 0.8~nm. 

Furthermore, we show, experimentally and by transfer matrix calculations, that the deposited graphene layers can easily be identified by optical microscopy, which greatly facilitates preparation and inspection. As the exfoliation technique can in principle be extended to ultra high vacuum conditions (see e.g.~\cite{Ritter08}) where preparing clean, atomically flat, crystalline substrates is a standard procedure, the graphene/substrate interface can be studied under very well controlled conditions. This could be used to study the influence of impurities and defects on the minimal conductivity in graphene \cite{Tan07,Chen08} in a more systematic way.

\section*{Acknowledgement}
We gratefully acknowledge financial support from the DFG in the frame of the SFB 616: {\em Energy dissipation at surfaces}. We thank V. Buck for granting us access to the Raman spectrometer and M. Neubert for his help with the measurements.

\newpage

\newpage
\begin{thebibliography}{1}

\bibitem{Novoselov05}
K.S. Novoselov, D. Jiang, F. Schedin, T.J. Booth, V.V. Khotkevich, S.V. Morozov, A.K. Geim, PNAS \textbf{102} (2005) 10451

\bibitem{Geim07}
A. Geim, K.S. Novoselov, Nature Materials {\bf 6} (2007) 183

\bibitem{Ni07}
Z.H. Ni, H.M. Wang, J. Kasim, H.M. Fan, T. Yu, Y.H. Wu, Y.P. Feng, Z.X. Shen, Nano Letters \textbf{7} (2007) 2758

\bibitem{Blake07}
P. Blake, E.W. Hill, A.H. Castro Neto, K.S. Novoselov, D. Jiang, R. Yang, T.J. Booth, A.K. Geim, Appl. Phys. Lett. \textbf{91} (2007) 063124

\bibitem{SAkcoeltekin09}
S. Akcöltekin, T. Roll, E. Akcöltekin, M. Klusmann, H. Lebius, M. Schleberger, Nucl. Instr. Methods B, in press.
\newblock doi:10.1016/j.nimb.2008.11.056    

\bibitem{Ferrari06}
A. C. Ferrari, J. C. Meyer, V. Scardaci, C. Casiraghi, M. Lazzeri, F. Mauri, S. Piscanec, D. Jiang, K. S. Novoselov, S. Roth, and A. K. Geim,   Phys. Rev. Lett. {\bf 97} (2006) 187401  

\bibitem{Abergel07}
D.S.L Abergel, A. Russell, V.I. Fal'ko,  Appl. Phys. Lett. {\bf 91} (2007) 063125  

\bibitem{Born}
M. Born, E. Wolf, Principals of Optics, Pergamon Press, 6th edition

\bibitem{CRC}
D.R. Lide ed., CRC Handbook of Chemistry and Physics, Taylor and Francis, 87th edition (1957)  

\bibitem{IGORPro}
Contrast values have been determined with the software package IGOR Pro by averaging over several linescans 

\bibitem{Obraztsova08}
E.A. Obraztsova, A.V. Osadchy, E.D. Obraztsova, S. Lefrant, I.V. Yaminsky, phys. stat. sol. (b) {\bf 245} (2008) 2055 

\bibitem{Adam07}
S. Adam, E.H. Hwang, V.M. Galitski, S.D. Sarma, PNAS {\bf 104} (2007) 18392

\bibitem{Chen08}
J.H. Chen, C. Jang, S. Adam, M.S. Fuhrer E. D. Williams and M. Ishigami, Nature Physics {\bf 4} (2008) 377

\bibitem{Du08}
X. Du, I. Skachako, A. Barker, E.Y. Andrei. Nature Nanotech. {\bf 3} (2008) 491

\bibitem{Stoeberl08}
U. Stöberl, U. Wurstbauer, W. Wegscheider, D. Weiss, J. Eroms, Appl. Phys. Lett. {\bf 93} (2008) 051906 

\bibitem{Ishigami07}
M. Ishigami, J.H. Chen, W.G. Cullen, M.S. Fuhrer, Nano Letters {\bf 7} (2008) 1643 

\bibitem{Ritter08}
K.A. Ritter, J.W. Lyding, Nanotechnology {\bf 19} (2008) 015704 

\bibitem{Tan07}
Y.-W. Tan, Y. Zhang, K. Bolotin, Y. Zhao, S. Adam, E.H. Hwang, S. Das Sarma, H.L. Stormer, P. Kim Phys, Rev. Lett. {\bf 99} (2007) 246803 

\end{thebibliography}
\end{document}